# Online Soft Error Tolerance in ReRAM Crossbars for Deep Learning Accelerators

Benyamin Khezeli
*Department of Computer Engineering*
*Amirkabir University of Technology*
*(Tehran Polytechnic)*
Tehran, Iran
ben.khezli@aut.ac.ir

Hamid R. Zarandi
*Department of Computer Engineering*
*Amirkabir University of Technology*
*(Tehran Polytechnic)*
Tehran, Iran
h_zarandi@aut.ac.ir

Elham Cheshmikhani
*Department of Computer Science and Engineering*
*Shahid Beheshti University*
Tehran, Iran
e_cheshmikhani@sbu.ac.ir

*Abstract*— Resistive Random-Access Memory (ReRAM) crossbar arrays are promising candidates for in-situ matrix-vector multiplication (MVM), a frequent operation in Deep Learning algorithms. Despite their advantages, these emerging non-volatile memories are susceptible to errors due to non-idealities such as immature fabrication processes and runtime errors, which lead to accuracy degradation in Processing-in-Memory (PIM) accelerators. This paper proposes an online soft error detection and correction method in ReRAM crossbar arrays. We utilize a test input vector and Error Correcting Codes (ECCs) to detect and correct faulty columns. The proposed approach demonstrates near fault-free accuracy for Neural Networks (NNs) on *MNIST* and *CIFAR-10* datasets, with low area overhead and power consumption compared to recent methods.

## I. Introduction

In recent years, the demand for efficient architectures to process neural network (NN) algorithms has surged due to their widespread application in various domains, from mobile devices to autonomous vehicles [6]. Traditional von Neumann architectures, where the processor and memory are separate entities, suffer from significant performance bottlenecks. This architecture requires extensive data movement between the processor and memory, leading to what is known as the memory wall problem [1], [3]. The separation of computation unit and memory results in substantial data transportation latency and increased power consumption, which are critical limitations for modern machine learning tasks [2]. Matrix-Vector Multiplication (MVM), a core operation in both the inference and training phases of neural networks is particularly affected by these inefficiencies. As Deep Neural Networks (DNNs) and particularly Convolutional Neural Networks (CNNs), become more prevalent, the inefficiencies of the von Neumann architecture become more pronounced [4]. Consequently, the gap between the energy and performance efficiency of today's hardware and the needs of these applications continue to widen, highlighting the urgent need for innovative architectural solutions like Processing-in-Memory (PIM) or Compute-in-Memory (CIM) paradigms, which integrate computation directly within memory arrays to mitigate these bottlenecks and enhance the overall system performance [5].

Emerging non-volatile memory technologies such as Spin-Transfer Torque Magnetic Random-Access Memory (STT-MRAM), Phase-Change Memory (PCM), and Resistive Random-Access Memory (ReRAM), are increasingly valued for their dual capabilities in storage and computation [7], [39], [40], [41]. Among these technologies, ReRAM is notable for its speed, low power consumption, near-zero leakage energy , high density , and multi-bit storage capacity [8], [10]. ReRAM crossbar structure, which inherently leverages parallelism for in-situ analog MVM makes it ideal for accelerating neural network applications. The ReRAM crossbar structure reduces the time complexity of MVM from $O(n^2)$ to $O(1)$, where $n$ refers to the matrix size, thus enhancing computational efficiency and energy saving [9].

Despite its advantages, ReRAM is prone to faults due to its relatively immature fabrication process and limited endurance cycles. These faults can be broadly classified into hard faults and soft faults. Hard faults, such as stuck-at-0 (SA0) or stuck-at-1 (SA1), occur when a cell's resistance becomes fixed and unchangeable, making the cell unusable. In contrast, soft faults are caused by variations in the target resistance due to factors like fabrication variations, programming noise, and read disturbances [14], [23]. Unlike hard faults, where the resistance is permanently fixed, the resistance in soft faults can still be adjusted, but these faults occur more randomly and frequently than hard faults, posing significant challenges for maintaining computational accuracy in critical applications [11].

In a ReRAM-Based Computing System (RCS), the impact of soft errors is compounded because unintended changes in ReRAM cell values affect the outcome of MVM operations, making it difficult to isolate and correct errors. This necessitates robust fault tolerance mechanisms to ensure reliable and accurate computation in deep learning accelerators.

Numerous previous works have been proposed methods to address the non-idealities in ReRAM-based systems. These methods can be categorized based on their suitability for offline versus online fault detection and correction. Some methods, test each cell sequentially using a specified test pattern order and detects faulty cells from a six-bit signature [10]. Another method accelerates testing by examining a group of adjacent cells simultaneously, based on the sneak path mechanism in ReRAM crossbars, which is suitable for offline testing due to its high time complexity [15]. However, these methods are impractical for online testing. Research indicates that even ReRAM chips that pass manufacturing tests often encounter faults during read and write operations due to limited write endurance [16], [17]. Methods such as [18], [19], [20], and [21] work with the assumption that fault distribution is known, which is inefficient for online error detection and correction; for example, the method in [21] targets stuck-at faults using voltage comparison and neuron reordering to mitigate the impact of hard errors. To mitigate accuracy degradation, several online methods such as [22], [24], and [25] have been proposed to operate at runtime. In [22], both hard and soft errors are addressed using redundant columns for weighted and non-weighted checksums, along with multiple test vectors. This method requires extra cycles for the test inputs and multiple redundant columns, making it both time and hardware-intensive. The method presented in [24] proposes partial checksum selection and computation schemes using majority voting-based checksum and hamming code-based checksum. While this technique corrects errors in a single column, it lacks the capability to correct multi-column errors. Another method presents an online block error correction technique combining checksum and hamming code-based linear coding to correct errors in a block of multiple columns. However, it cannot tolerate errors in multiple columns across different blocks and has a large decoding overhead, resulting in high power consumption [25]. Additionally, previous studies have been categorized based on targeted faults, with works such as [21], [26], [18], [20], and [27] focusing on stuck-at-fault tolerance, while a study such as [22] addresses soft and hard errors.

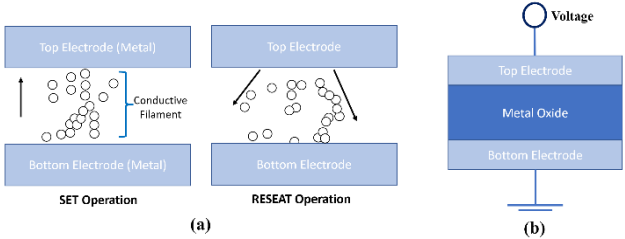

Fig. 1. ReRAM cell. (a) ReRAM operations and (b) ReRAM cell structure

In this paper, we propose a soft error-tolerant method to mitigate the impact of soft errors on DNN accuracy. The proposed method, considering that the weights of the pre-trained model are known and subsequently the conductance values of the ReRAM crossbars are also determined, uses a test input vector equal to the maximum applicable input voltage to detect faulty columns. This approach can identify both soft errors and hard errors, and correct columns with soft errors by reprogramming. The remainder of the paper is organized as follows: In Section II, we explore the ReRAM cell structure, ReRAM fault models, its crossbar architecture, how the MVM operation is performed in a ReRAM crossbar, and how the various layers of a CNN are mapped to the ReRAM crossbar. Section III describes the proposed method. Section IV presents the simulation results and compares the proposed method with recent works. Finally, Section V provides the conclusion of the paper.

## II. Preliminaries and Background

### A. ReRAM Cell Structure

Resistive Random-Access Memory known as ReRAM is a non-volatile memory technology that stores data by altering the resistance of passive two-terminal cells. These cells consist of a Metal-Insulator-Metal (MIM) structure, where an oxide layer is sandwiched between two metal layers. The resistance of the ReRAM cell changes between a High-Resistance State (HRS) and a Low-Resistance State (LRS), corresponding to the binary values 0 and 1, respectively. Fig. 1 shows the memory operations that occur by forming and rupturing conductive filament within the dielectric layer: applying a set voltage forms the filament (SET operation), switching the cell to the LRS, while reversing the polarity or modifying the voltage to reset voltage breaks the filament (RESET operation), returning the cell to the HRS [27].

Initially, a high voltage is required for the forming process to create the filament in a new ReRAM cell for the first time [28]. ReRAM technology supports both binary and Multiple-Level Cell (MLC) storage, where the resistance can be tuned to intermediate levels between HRS and LRS, enabling multiple bits of data per cell [31]. This scalability and versatility make ReRAM suitable for non-volatile memory design, digital and analog programmable systems, and neuromorphic computing structures. Its high density, low power consumption, and near-zero leakage energy make ReRAM ideal for main memory in applications like the Internet of Things (IoT), wearable devices, and biomedical devices, where power efficiency is critical [29], [30]. However, ReRAM faces significant reliability challenges, particularly concerning endurance and retention. Endurance refers to the number of write and erase cycles a ReRAM cell can undergo before it fails, with typical ReRAM cells enduring up to $10^6$ cycles, although some have been reported to last up to $10^9$ cycles [32]. Retention, on the other hand, is the ability of the ReRAM cell to maintain its resistance state over time without power. Some ReRAM cells exhibit rapid retention loss shortly after programming, which can compromise data stability [33]. These issues of endurance and retention variability are critical areas of ongoing research to improve the reliability of ReRAM technology.

### B. ReRAM Fault Models

ReRAM is susceptible to various faults, which can be broadly classified into hard and soft faults. Hard faults are due to defects occurring during the manufacturing process, wear-out, and endurance limitations, making the cell unusable [12], [13]. The best fault model to represent this type of fault is the stuck-at fault. These faults occur when a cell's resistance becomes fixed at a state. Examples include SA0 and SA1 faults, where the resistance state is fixed at HRS and LRS, respectively. These faults make the cell unusable and significantly impact the memory's reliability.

In contrast, soft faults arise from variations in the target resistance due to fabrication variations, temporary write failures, non-ideal write operations [14], programming noise, read disturbances, unbalanced programming pulses, parameter deviations, retention failures [23], and the diffusion of oxygen vacancies [12]. Unlike hard faults, the resistance in soft faults is not permanently fixed and can be adjusted. However, these faults occur randomly and more frequently, making them difficult to predict and manage.

Soft faults in ReRAM cells cause values to vary from the desired levels, leading to incorrect data that manifests as soft errors. These errors are particularly concerning in safety-critical applications where reliability is paramount. For instance, [11] demonstrated that a soft error in a Deep Neural Network (DNN) system could result in critical misclassifications, such as mistaking an approaching truck for a bird. This unpredictability poses a greater challenge for ensuring the integrity of Artificial Neural Networks (ANNs) computations.

Faults in ReRAM cells can also be categorized based on their occurrence and nature. Static faults originate during the fabrication process, resulting from inherent defects in manufacturing. On the other hand, dynamic faults typically occur during read and write operations in ReRAM cells that have passed initial fabrication tests. Dynamic faults arise from the operational environment and usage over time. Non-idealities such as non-linear voltage-current (V-I) characteristics and non-linear resistance levels can lead to deviations in resistance.

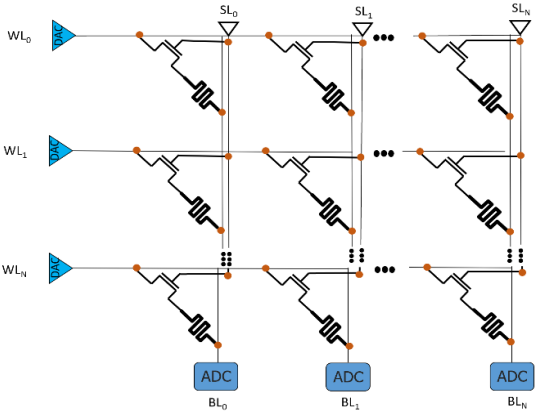

Fig. 2. 1T1R crossbar structure

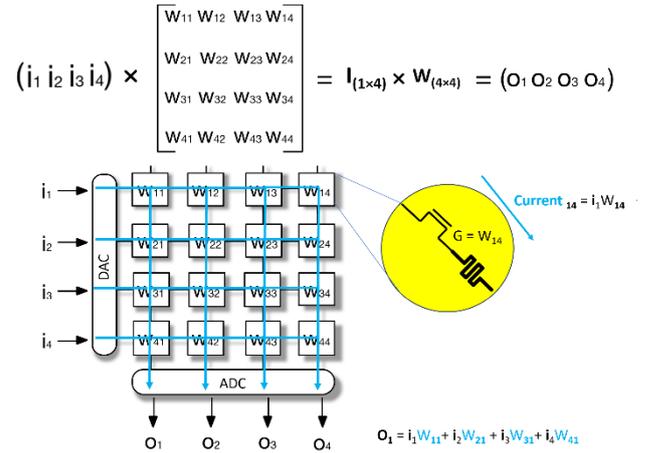

Fig. 3. Matrix-vector multiplication (MVM) in ReRAM crossbar

C. ReRAM Crossbar Structure

ReRAM crossbars are primarily designed in two configurations: the One-Resistor (1R) and the One-Transistor-One-Resistor (1T1R) structures. In the 1R configuration, each single resistor is accessed through Word Lines (WLs) and Bit Lines (BLs). This design enables the creation of dense memory arrays but suffers from the problem of unintended current paths, known as sneak-path currents, which can negatively affect data integrity [15]. In contrast, as shown in Fig. 2, the 1T1R structure includes an access transistor in each cell along with the resistor. This addition helps mitigate the sneak-path issue, offering improved reliability and control during data operations. Although slightly more complex, the 1T1R design is particularly suited for MVM tasks, which are crucial for high-performance computing applications [34], [8].

D. ReRAM Crossbar as MVM Accelerator in CNN

The ReRAM crossbar structure can accelerate MVM operations, reducing execution time from $O(n^2)$ to $O(1)$. MVM occurs on the ReRAM crossbar by programming the matrix values as conductance on the crossbar, while the input vector is applied as input voltages. In each ReRAM cell, according to Ohm's law ($I = \frac{V}{R} = V \times G$), a current is generated. According to Kirchhoff's current law, the currents in each column (bitline) are summed, which is equivalent to the dot product of the input vector and the corresponding column. Fig. 3 shows the multiplication of the input vector $I$ by the matrix $W$ on the ReRAM crossbar structure. The result of the MVM, which equals (O1, O2, O3, O4), corresponds to the output currents in each bitline column. The input and output vectors are digital, while the MVM operation occurs in the analog domain, necessitating the use of DAC and ADC converters. In the context of PIM-based accelerators utilizing the ReRAM crossbar for CNNs, as shown in Fig. 4 we can map two types of layers into the crossbar. One of them is the convolutional layer, where the feature maps are applied as inputs to the crossbar, and the kernels are mapped into the crossbar's conductance. For linear layers, the weight matrix is mapped as conductance, and the input matrix is applied as input voltages to the crossbar [35].

III. The Proposed Method

To overcome the mentioned errors, we propose a new method. This proposed method can be applied to any application where the values on ReRAM crossbars are predetermined, mapped once into the crossbar, and used multiple times (such as neural network weights that are programmed once and used repeatedly during inference). The proposed method works as follows: In neural network applications, the pre-trained weight values determine the conductance values in each crossbar after the mapping algorithm. By knowing these conductance values and applying a test input vector equivalent to the maximum applicable input voltage, we can obtain the equivalent current summation in each bitline. Using an $n$-bit precision ADC, we calculate the $n$-bit output of each column or bitline and store a portion of these bits when programming each crossbar. At runtime, the test vector is applied before the input vector during each MVM operation and is compared with the pre-stored values.

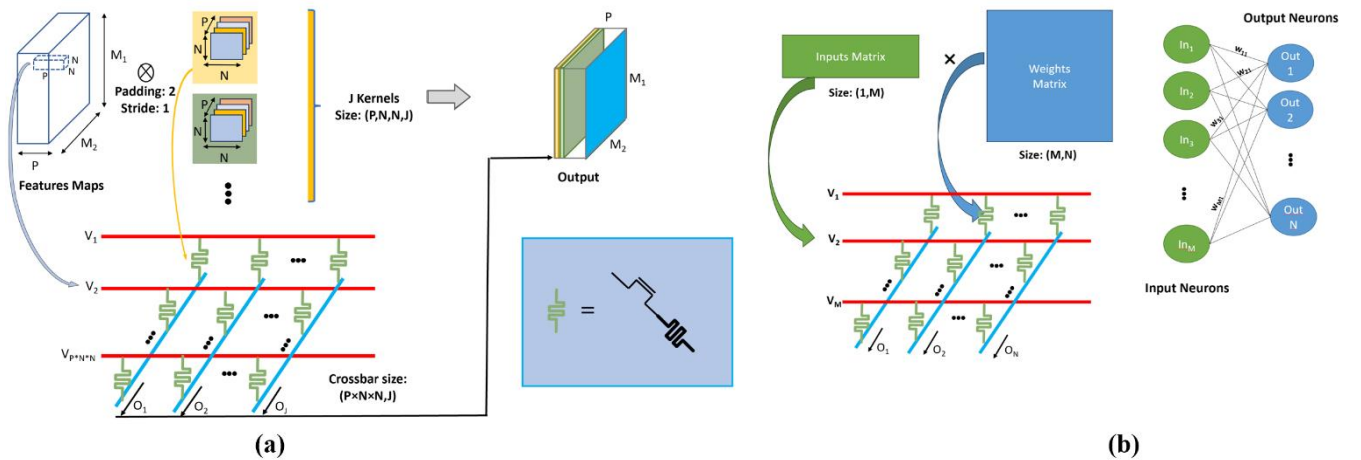

Fig. 4. Weight mapping based on ReRAM technology. (a) Convolution layer mapping; (b) Linear layer mapping

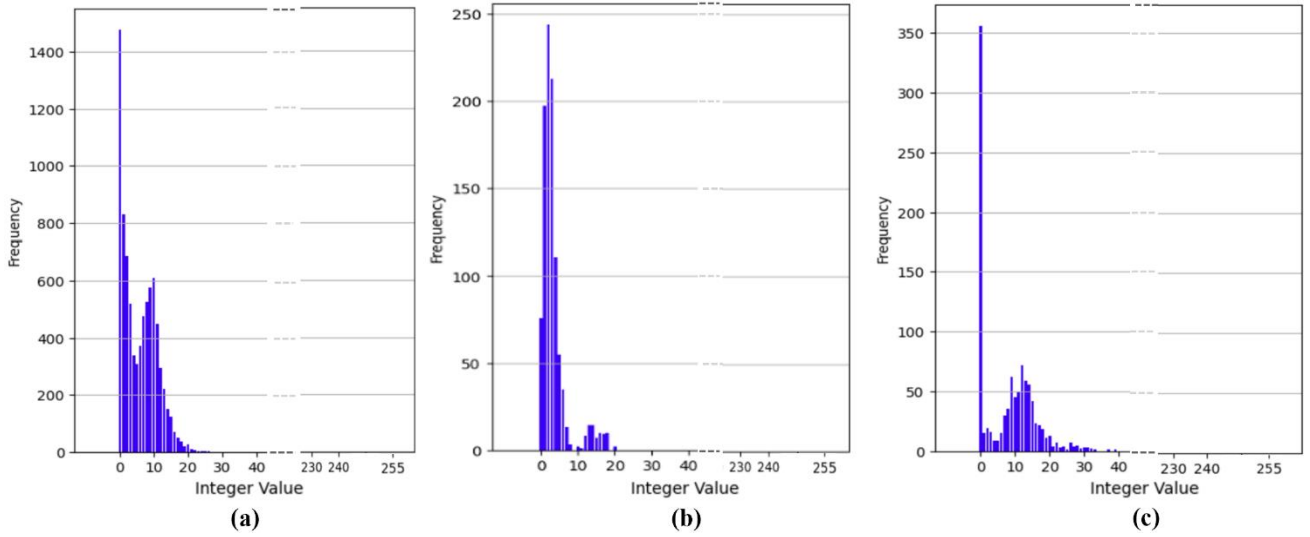

Fig. 5. ADC 8-bit output frequency on three NN models. (a) *SimpleNet*, (b) *LeNet5* and (c) *ResNet18*

Fig. 5 shows the frequency of the 8-bit ADC output for three neural network models with $128 \times 128$ crossbars, 8-bit resolution ADCs at each bitline output, and a maximum input voltage of 0.3V. As evident, the output values are predominantly in the range of smaller values, resulting in more frequent changes in the Least Significant Bits (LSBs). Therefore, by storing a few LSBs that result from the MVM operation of the maximum input voltage and fault-free crossbar's conductance, we can achieve an accuracy close to the fault-free model.

Fig. 6 shows the implementation of this method, assuming the use of $128 \times 128$ crossbars and 8-bit resolution ADCs with the maximum input voltage of 0.3V. in this method, in the additional cycle before the input vector is applied to the ReRAM crossbar, the test input vector is applied. By performing the in-situ analog MVM operation that takes place in the crossbar, the digitized outputs are calculated at the ADC output. By comparing the generated 4 LSBs during the application of the test input vector with the initially stored 4-bit redundancy, we can detect a faulty column. For this, we need a 4-bit comparator circuit that compares the calculated 4 LSBs with the pre-stored 4-bit value in each column. If the two 4-bit values are the same, no error is detected. If they are not the same, the conductance values of the targeted column are reprogrammed, and the input vector is applied in the next

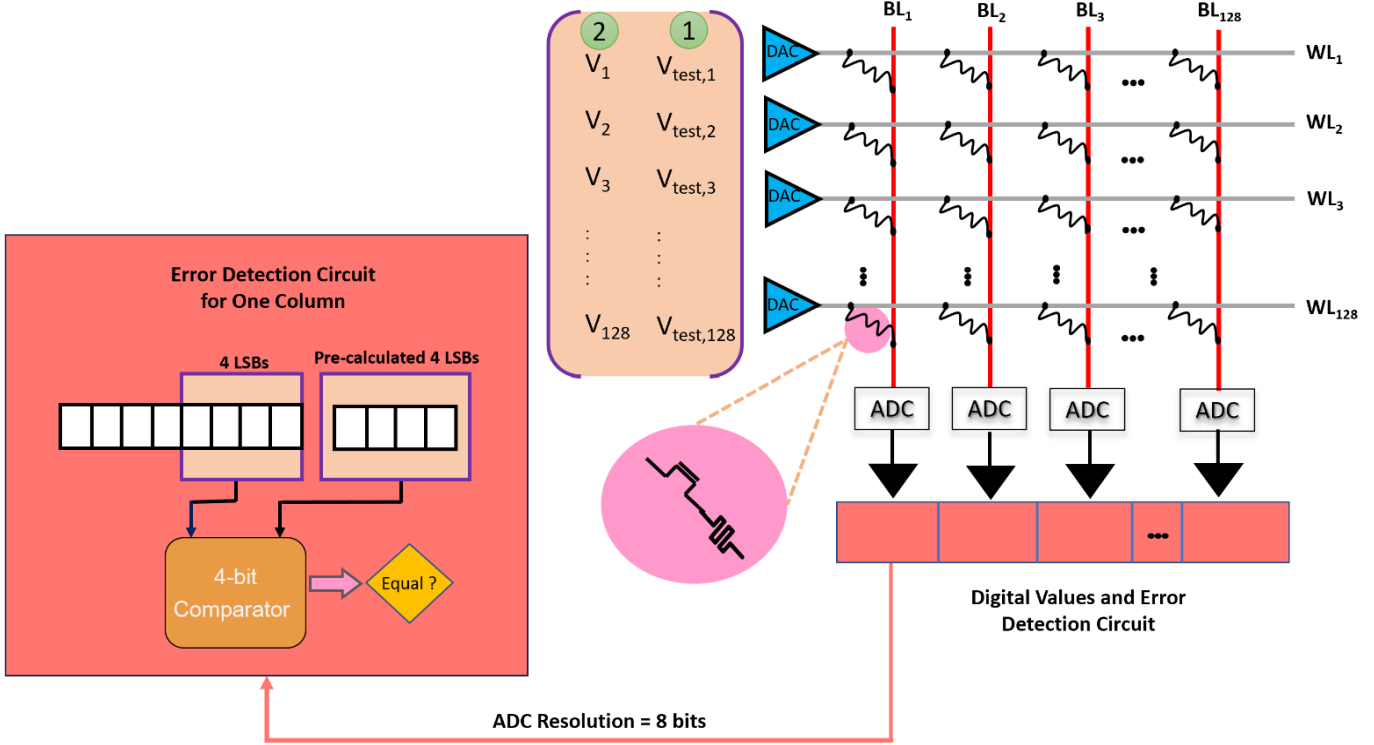

Fig. 6. The proposed method using 4 LSBs as redundancy

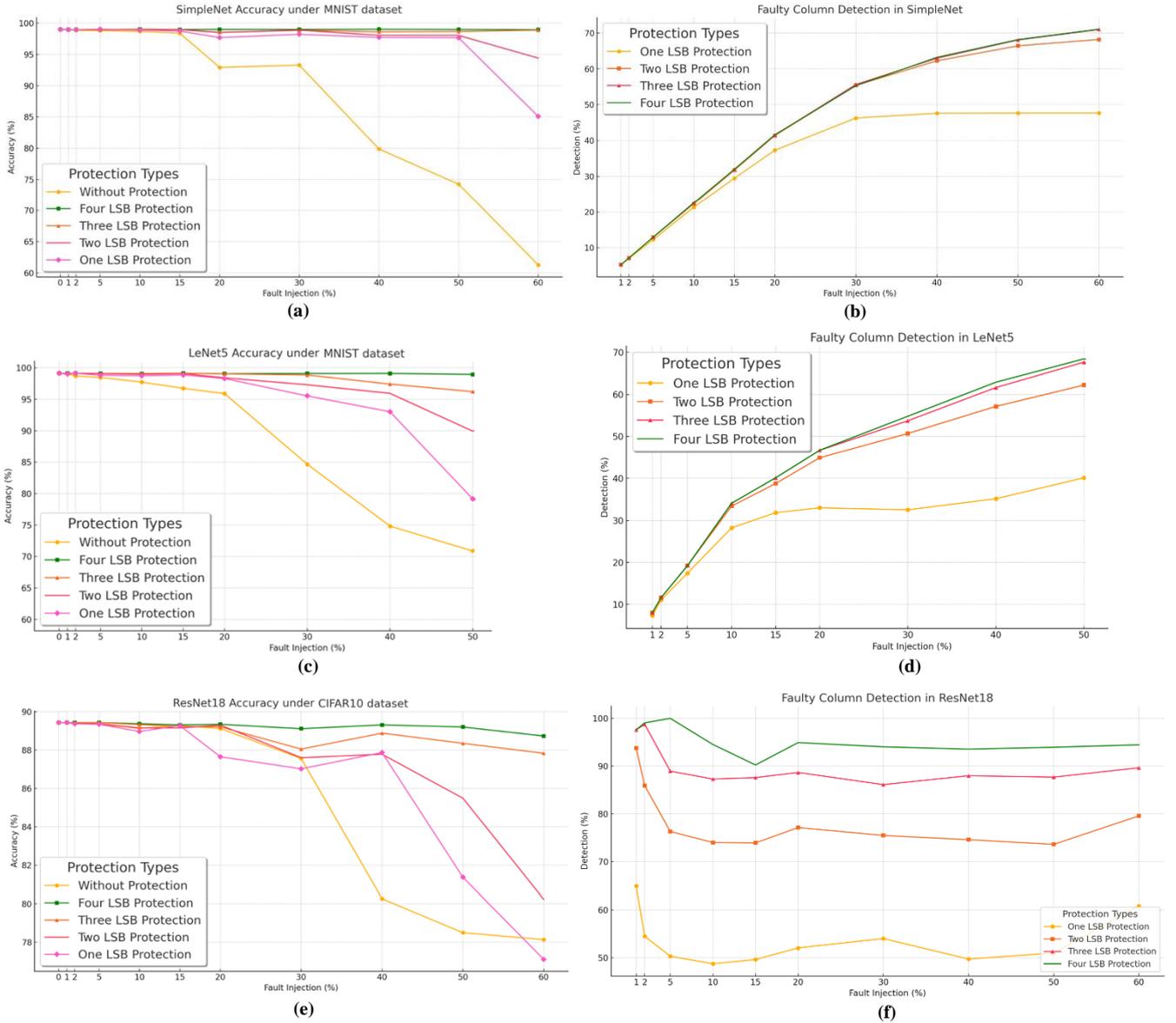

Fig. 7. Accuracy and fault detection of three models in the presence of different percentages of faults. (a) *SimpleNet* accuracy, (b) *SimpleNet* fault detection, (c) *LeNet5* accuracy, (d) *LeNet5* fault detection, (e) *ResNet18* accuracy and (f) *ResNet18* fault detection

step. The reason we apply the maximum input voltage is that if cells in a column have had their conductance altered due to a soft error, this change will be multiplied by a larger number. Consequently, it is more likely that the output value of the ADC differs with the ADC output in the fault-free scenario.

## IV. Simulation and Evaluation

To evaluate the proposed method, we analyze the performance of three neural network architectures with and without error detection, injecting faults at various percentages. The architectures employed are *SimpleNet*, *LeNet5*, and *ResNet18*, based on the structures and parameters in [36]. The accuracy of *SimpleNet* and *LeNet5* models is tested on the *MNIST*, while the *ResNet18* model is tested on the *CIFAR-10*.

For simulating the behavior of the PIM-based accelerators with ReRAM crossbar structures, we utilized the *MemTorch* simulation framework [37]. In this framework, based on the CNN accelerator model used in [35], we employ the following configuration: crossbars are 128×128, all ADCs have a resolution of 8 bits, and the maximum input voltage is set to +0.3V, as specified by the simulator.

The linear layers of the aforementioned neural networks are mapped and accelerated on the crossbar. Then, their accuracy is evaluated in both fault-free and faulty models. Subsequently, the accuracy of the models using the proposed method was tested under varying fault percentages. Alongside accuracy, the fault detection rate for each case is also reported.

The proposed method operates by considering 1, 2, 3, and 4 LSBs as redundancy and evaluate the neural network accuracy in the presence of different fault percentages. As shown in Fig. 7, by considering 4 LSBs as redundancy alongside the main data, the neural network accuracy closely approaches the fault-free model. Moreover, this method demonstrates higher accuracy and fault detection rates under high fault rates compared to other methods, enhancing the system's fault tolerance.

Table 1. Error detection and correction overhead for the proposed method

| Matrix size | Majority Voting based Checksum Scheme [24] | | | Hamming Code based Checksum Scheme [24] | | | Fixed Block ECC (7,4) [25] | | | The Proposed Method | | |
|---|---|---|---|---|---|---|---|---|---|---|---|---|
| | Area ($\mu m^2$) | Latency (ns) | Power (mW) | Area ($\mu m^2$) | Lateny (ns) | Power (mW) | Area ($\mu m^2$) | Latency (ns) | Power (mW) | Area ($\mu m^2$) | Latency (ns) | Power (mW) |
| 64×64 | 7987.96 | 1.61 | 8.66 | 6767.78 | 2.44 | 5.75 | 14971 | 6.33 | 7.44 | 539.49 | 20.09 | 0.17 |
| 128×128 | 15829.96 | 2.18 | 18.36 | 13953.70 | 2.96 | 11.77 | 29593 | 6.95 | 15.01 | 2160.71 | 20.15 | 0.67 |
| 256×256 | 31210.33 | 2.31 | 37.92 | 27419.79 | 3.30 | 23.35 | 59187 | 7.53 | 30.20 | 8648.34 | 20.24 | 2.67 |

Additionally, the proposed method was compared with recent state-of-the-art methods [24], and [25] in terms of area, delay, and power using the *NVSim* [38] and *Synopsys HSPICE circuit* simulation frameworks, based on 32nm technology. The *NVSim* tool was utilized to simulate the reprogramming of faulty columns and to analyze power and delay, while the *Synopsys HSPICE* framework was used to model the area, delay, and power of the comparator. Table 1 presents the simulation results, demonstrating that the proposed method gain significantly lower area and power consumption compared to [24], and [25], with a negligible increase in delay due to the application of the test input vector and the reprogramming of the crossbar.

## V. Conclusion

In this paper, we presented an efficient method for online soft error detection and correction in ReRAM crossbar arrays used in deep learning accelerators. By leveraging pre-known neural network weights and corresponding conductance values, our approach employs a test input vector to detect both soft and hard errors, and specially reprogram faulty columns to correct soft errors. Our method demonstrated near fault-free accuracy on the *MNIST* and *CIFAR-10* datasets, while maintaining low area overhead and power consumption. We evaluated three neural network architectures: *SimpleNet*, *LeNet5*, and *ResNet18*, and found that considering 4 LSBs as redundancy allowed the neural networks to achieve near fault-free accuracy, even under high fault rates. Comparative analysis using *NVSim* and *Synopsys HSPICE* circuit simulation frameworks showed that our approach significantly reduces area and power consumption, albeit with an increase in delay due to the test input vector and reprogramming process. These results highlight the proposed method's effectiveness in enhancing the accuracy and reliability of ReRAM-based PIM accelerators in the presence of soft errors.